\documentclass[aps,preprint,groupedaddress,showpacs]{revtex4}

\def\vb#1{\mbox{\boldmath$#1$}}
\def\pd#1#2{\frac{\partial #1}{\partial #2}}

\def\wh#1{\widehat{#1}}
\def\bdot{\,\vb{\cdot}\,}
\def\btimes{\,\vb{\times}\,}

\def\bhat{\wh{{\sf b}}}

\def \bpar*{{ B}^{*}_{\parallel}}

\def\x{\mbox {\boldmath $\xi$}}
\def\c{{\bf C}}

\def\L{\mathcal L}

\def\K{\mathcal K}

\begin{document}

\title{Nonlocal Nonlinear Electrostatic Gyrofluid Equations: \\ A four-moment model}
\author{D.~Strintzi and B.D.~Scott}
\affiliation{Max-Planck-Institut f\"ur Plasmaphysik, EURATOM Association\\
                D-85748 Garching, Germany}
\author{A.J.~Brizard}
\affiliation{Department of Physics, Saint Michael's College \\
                One Winooski Park, Colchester, VT 05439, USA}

\date{October 8, 2004}

\begin{abstract}

Extending a previous single-temperature model, an electrostatic gyrofluid model that includes anisotropic temperatures $(T_{\|} \neq T_{\bot})$ and can treat general nonlinear situations is constructed. The model is based on a Lagrangian formulation of gyrofluid dynamics, which leads to an exact energy conservation law. Diamagnetic cancelations are inserted manually in such a way that energy conservation is preserved. Comparison with previous models shows a very good agreement for zero-Larmor-radius terms in the gyrofluid equations of motion.

\end{abstract}

\pacs{52.35.Ra, 52.30.Ex, 52.65.Tt}

\maketitle

\pagebreak

\vfill\eject

\section{Introduction}

The gyrofluid model \cite{brizard}-\cite{snyder} is one of the most commonly used tools for understanding and explaining tokamak turbulent transport. Although it is less accurate than the gyrokinetic approach, it is computationally more efficient and economic, and it 
retains most of the qualitative features necessary to explain turbulent transport \cite{dorland}-\cite{snyder}.

To correctly compute turbulent transport, and especially for those computations that treat large-amplitude disturbances, or those that treat strong spatial variations in the plasma parameters, a gyrofluid model should satisfy an exact energy conservation law. In linear theory, a small deviation from energy conservation leads to a small change in the growth rate. Turbulence, however, also includes several conservative transfer pathways that influence the overall statistical equilibrium. Violation of energy conservation in this case can lead to a source of free energy, and, hence, large errors in the description of the turbulent state. Moreover, to study the nonlinear evolution subsequent to a linear growth rate, the model should be fully nonlinear, while conserving energy.

In a previous paper \cite{ours}, we derived a fully nonlinear set of electrostatic gyrofluid equations that satisfy an exact energy theorem. This set contains the evolution equations for density, momentum, and perpendicular temperature, as well as the polarization equation for the electrostatic potential. The energy conservation law, which is guaranteed by the Noether method through the Lagrangian formulation of gyrofluid dynamics, is a unique and important feature of this model, which is thought to be especially important for the numerical simulation of turbulence. Although our previous single-temperature gyrofluid model is consistent, it does not contain the parallel-temperature effects necessary for a quantitatively accurate description of turbulence in magnetized plasmas. In the present paper, we extend our previous model to include these effects. Although the model is still electrostatic, it should give an accurate description of low $\beta$ turbulent plasmas.

The extension to anisotropic temperatures $(T_{\|} \neq T_{\bot})$ turns out to be nontrivial. In the single-temperature case, the constraints used in the variation of the Lagrangian density can be chosen to be the particle and entropy conservation laws. In the two-temperature case, one needs an additional constraint and one can no longer rely on (only) conservation laws. The approach used here is to derive the constraints for the parallel and perpendicular pressure through the full Chew-Goldberger-Low (CGL) equations \cite{CGL,weiland}. A further difference with our previous work is the use of the virtual fluid displacement $\x$ in deriving {\it Eulerian} variations for the gyrofluid moments, instead of the Lagrangian-multiplier method (which turns out to be algebraically more involved).

Gyrofluid equations that include anisotropic parallel and perpendicular pressures have been derived previously through the derivation of moments of the gyrokinetic equation \cite{brizard}-\cite{snyder}. Although these equations do not conserve energy, they agree to a large extent with our present gyrofluid model.

The remainder of this paper is organized as follows. In Sec.~II, we present the gyrofluid Lagrangian and give the constraints on the density and anisotropic pressures used in the variational principle. In Sec.~III, we present the variation of the gyrofluid action functional, and derive the gyrofluid equations of motion before the insertion of diamagnetic cancelations, as well as the polarization equation. In Sec.~IV, we derive the energy conservation law by applying the Noether method. In Sec.~V, we describe how the diamagnetic cancelations are inserted into our gyrofluid equations and present the final equations for density, momentum, parallel and perpendicular pressures. We also present the explicit form of the energy conservation law with each energy-exchange term clearly identified. Lastly, in Sec.~VI, we summarize our work and present our conclusions.

\section{Gyrofluid Lagrangian Density and Lagrangian \\ Constraints}

\subsection{Gyrofluid Lagrangian Density}

The four-moment gyrofluid Lagrangian for the anisotropic-temperature model is constructed in the same way as that of the one-temperature model 
\cite{ours}, based on the work of Pfirsch and Correa-Restrepo \cite{pfirsch}. Here, the gyrofluid Lagrangian density is defined as
\begin{equation}
\L_{f} \;=\; \frac{1}{2}\;nm\, \left( u_{\|}^{2} \;+\; |{\bf u}_E|^{2}\right) \;-\; \left( p_{\perp} \;+\; \frac{p_{\|}}{2} \right) \;+\; en \left(
{\bf A} \bdot \frac{{\bf u}}{c} \;-\; \|\phi\| \right),
\label{eq:a1}
\end{equation}
where ${\bf u}_E = (c/B)\,\bhat\btimes\nabla\phi$ is the $E\times B$ velocity, $u_{\|} \equiv {\bf u}\bdot\bhat$ denotes the gyrofluid velocity parallel to the magnetic field ${\bf B} \equiv \nabla\btimes{\bf A} = B\,\bhat$, and the 
{\it nonlocal} operator $\|...\|$ denotes a gyrofluid gyro-averaging operator defined as follows. First, we note that, in our four-moment model, the electrostatic gyro-potential $\|\phi\|$ depends on the perpendicular temperature $T_\perp \equiv p_{\bot}/n$ and is assumed not to depend on higher-order gyrofluid moments. Thus, a general expression for $\|\phi\|$ can be given as
\begin{equation}
\|\phi\| \;\equiv\; {\cal P}\left(\frac{\rho^{2}}{2}\;\nabla_{\bot}^{2}\right)\phi \;=\; \sum_{k}\;\frac{{\cal P}^{(k)}(0)}{2^{k}\,k!}\; \rho^{2k}
\nabla_{\bot}^{2k}\phi,
\label{eq:phi_gyro}
\end{equation}
where $\rho^{2} = (mc^{2}/e^{2}B^{2})\,T_{\bot}$ is the squared Larmor radius and ${\cal P}(x)$ represents a smooth function such that ${\cal P}(0) \equiv 1$, i.e., we have $\|\phi\| \rightarrow \phi$ in the zero-Larmor-radius (ZLR) limit. Two choices for ${\cal P}(x)$ will be discussed below; here, we note that the operator $\nabla_{\bot}^{2}$ in the Taylor expansion (\ref{eq:phi_gyro}) acts on $\phi$ alone. The Hermitian conjugate operator $\|...\|^{\dag}$ is defined in terms of Eq.~(\ref{eq:phi_gyro}) as
\begin{equation}
\|\chi\|^{\dag} \;\equiv\; \sum_{k}\; \frac{{\cal P}^{(k)}(0)}{2^{k}\,k!}\; \nabla_{\bot}^{2k}\left( \chi\,\rho^{2k}
\right) \;=\; {\cal P}\left(\nabla_{\bot}^{2}\;\frac{\rho^{2}}{2}\right)\;\chi,
\label{eq:dagger}
\end{equation}
where $\chi$ is an arbitrary function and we made use of the identity
\begin{eqnarray}
\chi\|\phi\| & = & \phi\;\|\chi\|^{\dag} \;+\; \nabla_{\bot}\bdot\left[\; \left( \chi_{1} \;+\; 
\nabla_{\bot}^{2}\chi_{2} \;+\; \cdots\right) \nabla_{\bot}\phi \;-\; \phi\;\nabla_{\bot}\left( \chi_{1} \;+\; \nabla_{\bot}^{2}\chi_{2} \;+\; \cdots\right) \right. \nonumber \\
 &   &\left.+\; \left( \chi_{2} \;+\; \cdots\right) \nabla_{\bot}\nabla_{\bot}^{2}\phi \;-\; \nabla_{\bot}^{2}\phi\;\nabla_{\bot}\left( \chi_{2} \;+\;
\cdots\right) \;+\; \cdots \;\right],
\label{eq:dag_identity}
\end{eqnarray}
where 
\[ \chi_{k} \;\equiv\; \pd{(\chi\|\phi\|)}{(\nabla_{\bot}^{2k}\phi)} \;=\; \frac{{\cal P}^{(k)}(0)}{2^{k}\,k!}\; 
\chi\,\rho^{2k}. \]
Note that the operator $\nabla_{\bot}^{2}$ in Eq.~(\ref{eq:dagger}) now acts on $\chi$ and powers of the Larmor radius 
$\rho$.

The $E\times B$ kinetic term $m |{\bf u}_E|^{2}/2$ in Eq.~(\ref{eq:a1}) is the long-wavelength form of {\it gyro-screening} (i.e., the ZLR part of the second-order potential term in the gyrokinetic Hamiltonian \cite{hahm}), which plays a major role in the derivation of the polarization equation. Although the $E\times B$ kinetic term is only an approximation of the full second-order gyrokinetic Hamiltonian, it is useful since the gyro-screening correction to the potential is important only for large flow velocities and long wavelengths. Thus, the addition of full finite-Larmor-radius (FLR) effects for this quadratic term will not alter any computational results. The fact that the gyro-screening term appears only in the Hamiltonian leads to the appearance of polarization effects only in Poisson's equation for the scalar potential $\phi$, which is desirable for computational reasons. 

Although the drift-fluid Lagrangian of Ref.~\cite{pfirsch} and our gyrofluid Lagrangian are similar, there exist three differences: (a) the gyrofluid Lagrangian (\ref{eq:a1}) contains FLR corrections to the scalar potential $\phi$, (b) the anisotropic temperatures $T_\perp$ and $T_{\parallel}$ are treated separately rather than as a single isotropic temperature $T$, and (c) we have arranged the terms such that the gyro-screening term appears only in the Hamiltonian (i.e., not in the {\it symplectic} part involving contraction with ${\bf u}$; see Ref.~\cite{ours} for further details). The total Lagrangian density for our four-moment gyrofluid model is the sum of the gyrofluid Lagrangian density (\ref{eq:a1}) and the Lagrangian density of the electric field, expressed as
\begin{equation}
\L \;=\; \frac{1}{2}\;nm\, \left( u_{\|}^{2} \;+\; |{\bf u}_E|^{2}\right) \;-\; n \left( T_{\perp} \;+\; \frac{T_{\|}}{2} \right) \;+\; en \left(
{\bf A} \bdot \frac{{\bf u}}{c} \;-\; \|\phi\| \right) \;+\; {|{\bf E}|^{2}\over 8\pi},
\label{eq:Lag_total}
\end{equation}
where the variational fields are the four gyrofluid moments $(n,{\bf u},p_{\|},p_{\bot})$ for each fluid species (here, summation over particle species
is assumed wherever appropriate) and the electrostatic potential $\phi$.

\subsection{Lagrangian Constraints}

In order to proceed with our variational principle, we introduce constraints on the gyrofluid moments 
$(n,p_{\|},p_{\bot})$, based on the continuity (mass conservation) equation
\begin{equation}
\frac{dn}{dt} \;+\; n\;\nabla\bdot{\bf u} \;=\; 0, \label{eq:continuity}
\end{equation}
where $d/dt = \partial/\partial t + {\bf u}\bdot\nabla$ denotes the total time derivative, and the modified CGL equations for the perpendicular and parallel pressures \cite{CGL,weiland}:
\begin{eqnarray}
\frac{dp_{\|}}{dt} \;+\; p_{\|}\;\nabla\bdot{\bf u} \;+\; 2\,p_{\parallel}\;\bhat\bhat:\nabla{\bf u} & = & -\;2\,\nabla\bdot
{\bf q}_{\parallel\perp} \;+\; 4\,{\bf q}_{\parallel \perp}\bdot (\bhat\bdot \nabla\bhat), \label{eq:p_parq} \\
\frac{dp_{\bot}}{dt} \;+\; p_{\bot}\;\nabla\bdot{\bf u} \;+\; p_{\perp}\;({\bf I} - \bhat\bhat):\nabla{\bf u} & = & -\;\nabla\bdot{\bf q}_{\perp\perp} 
\;-\; 2\,{\bf q}_{\parallel \perp}\bdot (\bhat\bdot \nabla\bhat), \label{eq:p_perpq}
\end{eqnarray}
where ${\bf q}_{\parallel \perp}$ and ${\bf q}_{\perp \perp}$ are the parallel and perpendicular heat fluxes, respectively. The pressure constraints, without the terms including the heat fluxes, can be also be obtained by general conservation laws. If the heat fluxes are removed from the pressure equations (\ref{eq:p_parq})-(\ref{eq:p_perpq}), all gyrofluid quantities $\eta^{a} = (n,p_{\|},p_{\bot})$ are directly related to the velocity ${\bf u}$. 
Using these equations as constraints for the Lagrangian density (\ref{eq:Lag_total}), we can obtain the evolution equation for the gyrofluid velocity 
${\bf u}$. 

We obtain expressions for the Lagrangian variations $\Delta\eta^{a}$ in terms of the virtual fluid displacement $\x$ by taking the limits 
\[ \lim_{\Delta t \rightarrow 0}\,\left( \frac{d\eta^{a}}{dt}\;\Delta t \right) \;\equiv\; \Delta\eta^{a} \;\;\;{\rm and}\;\;\;
\lim_{\Delta t \rightarrow 0}\,\left( {\bf u}\;\Delta t \right) \;\equiv\; \x, \]
so that the variations of the pressures from Eqs.~(\ref{eq:p_parq}) and (\ref{eq:p_perpq}) become
\begin{equation}
\left. \begin{array}{rcl}
\Delta p_{\|} & = & -\,p_{\|}\;\nabla\bdot\x \;-\; 2\,p_{\|}\;\bhat\bhat:\nabla\x \\
 &  & \\
\Delta p_{\bot} & = & -\,p_{\bot}\;\nabla\bdot\x \;-\; p_{\bot}\;({\bf I} - \bhat\bhat):\nabla\x 
\end{array} \right\}.
\label{eq:Lag_constrap}
\end{equation}
The heat-flux terms in Eqs.~(\ref{eq:p_parq})-(\ref{eq:p_perpq}), which were not used in the Lagrangian variations 
(\ref{eq:Lag_constra}) for $p_{\|}$ and $p_{\bot}$, are added to the pressure equations later on since they play an important role in the diamagnetic cancelations; these cancelations refer to the fact that the diamagnetic velocity does not lead to advection. The present Lagrangian formulation also requires one additional Lagrangian variation: 
\begin{equation}
\Delta{\bf u} \equiv  d\x/dt \;=\; \partial\x/\partial t + {\bf u}\bdot\nabla\x
\end{equation}
for the fluid velocity ${\bf u}$. For a correct treatment of FLR effects associated with the gyrofluid electrostatic potential $\|\phi\|$, instead of the
pressure variations we will use the temperature variations, such that the
set of the Lagrangian variations used will be:
\begin{equation}
\left. \begin{array}{rcl}
\Delta n & = & -\,n\;\nabla\bdot\x \\
 &  & \\
\Delta{\bf u} \; & = & \; \partial\x/\partial t + {\bf u}\bdot\nabla\x \\
 &  & \\
\Delta T_{\parallel}\; & = &  \; -\, 2T_{\parallel} \; \bhat\bhat :\nabla \x \\
 &  & \\
\Delta T_{\bot} \; & = & \; -\,T_{\bot}\;\left({\bf I} \;-\; \bhat\bhat\right):\nabla\x
\end{array} \right\}.
\label{eq:Lag_constra}
\end{equation}
where $\Delta n$ is obtained from Eq. (\ref{eq:continuity}).

Lastly, one should note that the gyrofluid Lagrangian density (\ref{eq:Lag_total}) was constructed, not derived, and that the method adopted here is not a mathematical derivation of the gyrofluid Lagrangian from the gyrokinetic (single-particle) Lagrangian. The validity of the present gyrofluid Lagrangian follows from the validity of the resulting gyrofluid equations, as well as the energy conservation law. The same applies to the constraint equations, e.g.,
Eqs.~(\ref{eq:p_parq})-(\ref{eq:p_perpq}), which can be introduced arbitrarily, and their validity follows from the validity of the resulting evolution equations.

\section{Gyrofluid Dynamical Equations}

In this Section, we present a variational principle based on the action functional $S \equiv \int \L\, d^{3}x\,
dt$ using the virtual-displacement method associated with the Eulerian variations $(\delta n, \delta{\bf u}, 
\delta T_{\|}, \delta T_{\bot})$. Here, each Eulerian variation $\delta\chi$ is defined in terms of its Lagrangian variation $\Delta\chi$ as $\delta\chi \equiv \Delta\chi - \x\bdot\nabla\chi$, so that, using 
Eq.~(\ref{eq:Lag_constra}), we find
\begin{equation}
\left. \begin{array}{rcl}
\delta n & = & -\;\nabla \bdot (n\;\x) \\
 &  & \\
\delta {\bf u} & = & \partial_{t}\x \;+\; ({\bf u}\bdot \nabla )\x \;-\; (\x \bdot \nabla){\bf u} \\
 &  & \\
\delta T_{\|} & = & -\x \;\bdot \nabla T_{\|} \;-\; 2T_{\|}\; \bhat\bhat:\nabla\x \\
 &  & \\
\delta T_{\bot} & = & -\;\nabla \bdot(T_{\bot}\;\x ) \;+\; T_{\bot}\;\bhat\bhat:\nabla\x
\end{array} \right\}.
\label{eq:constraints}
\end{equation}
From the Eulerian variation of the gyrofluid Lagrangian (\ref{eq:Lag_total}), we can now derive the moment equation for the gyrofluid velocity ${\bf u}$, as well as the polarization equation for the scalar potential $\phi$. Note that the Eulerian variation $\delta T_{\bot}$ in Eq.~(\ref{eq:constraints}) is used also in connection with the FLR dependence of $\|\phi\|$:
\begin{equation}
\delta\|\phi\|(\phi, T_{\bot}) \;\equiv\; \|\delta\phi\| \;+\; \delta T_{\bot}\; \pd{\|\phi\|}{T_{\bot}},
\label{eq:delta_phi}
\end{equation}
where
\begin{equation} 
\pd{\|\phi\|}{T_{\bot}} \;=\; \frac{mc^{2}}{2\,e^{2}B^{2}}\;{\cal P}^{\prime}\left(\frac{\rho^{2}}{2}\;\nabla_{\bot}^{2}\right)\;\nabla_{\bot}^{2}\phi \;\equiv\; \Omega(\phi, T_{\bot}; B),
\label{eq:Omega_def}
\end{equation}
and ${\cal P}^{\prime}(x)$ denotes the first derivative of ${\cal P}(x)$. As a result of the mass scaling in 
Eq.~(\ref{eq:Omega_def}), the ion FLR correction $\Omega_{i}$ is much larger than the electron FLR correction $\Omega_{e}$.

\subsection{Variation of the Lagrangian Density}

It is straightforward to derive the total variation $\delta\L$ of the gyrofluid Lagrangian density (\ref{eq:Lag_total}), which depends on the variational fields $(n,{\bf u},T_{\parallel},T_{\perp},\phi)$, so that we obtain
\begin{equation}
\delta\L \;=\; \left( \nabla\delta\phi\;\bdot\pd{\L}{(\nabla\phi)} \;-\; en\;\|\delta\phi\| \right) \;+\; \delta n\;
\pd{{\cal L}}{n} \;+\; \delta{\bf u}\bdot\pd{\L}{{\bf u}} \;-\; \left[\; \frac{n}{2}\,\delta T_{\|} \;+\; (1 +
 e\,\Omega)\;n\,\delta T_{\bot} \;\right].
\label{eq:Lag_var}
\end{equation}
Here, from Eq.~(\ref{eq:Lag_total}), we find
\begin{equation} 
\pd{\L}{n} \;=\; \frac{m}{2}\;u_{\|}^{2} \;+\; \frac{e}{c}\;{\bf A}\bdot{\bf u} \;-\; e\;\psi \;-\;\left(T_{\perp}+{1\over 2}\, T_{\parallel} \right) \;\;\;{\rm and}\;\;\;
\pd{\L}{{\bf u}} \;=\; n \left( m\;u_{\|}\,\bhat \;+\; \frac{e}{c}\;{\bf A} \right),
\label{eq:Lag_deriv}
\end{equation}
where we introduced the definition 
\begin{equation} 
e\,\psi(\phi, T_{\bot}; B) \;\equiv\; e\,\|\phi\| \;-\; (mc^{2}/2\,B^{2})\;|\nabla\phi|^{2},
\label{eq:psi_def}
\end{equation}
which combines the linear (with FLR corrections) and nonlinear (in the ZLR limit) electrostatic terms of the gyrokinetic Hamiltonian \cite{hahm}. 

By inserting the Eulerian variations (\ref{eq:constraints}) into the gyrofluid Lagrangian variation (\ref{eq:Lag_var}), we obtain (after some algebra)
\begin{eqnarray}
\delta\L & = & -\; \x \bdot\left\{\; \pd{}{t} \pd{\L}{{\bf u}} \;+\; \nabla\bdot\left({\bf u}\;\pd{\L}{{\bf u}} \right) 
\;+\; \nabla{\bf u}\bdot\pd{\L}{{\bf u}} \right. \nonumber \\
&   & \left. \;+\; \nabla\bdot{\sf P}^{*} \;-\; n \left[ \nabla\left( \pd{{\cal L}}{n}\;+\; T_{\perp} \;+\; {1\over 2}\, T_{\parallel} \right) \;+\; 
e\,\Omega\;\nabla T_{\bot} \right]  \;\right\} \nonumber \\
 &  &\mbox{}-\;\delta\phi \left( \nabla\bdot\pd{\L}{(\nabla\phi)} \;+\; \|en\|^{\dag} \right) \;+\; \pd{\Lambda}{t} \;+\; \nabla\bdot\vb{\Gamma},
\label{eq:Eul_Lag}
\end{eqnarray}
where the tensor ${\sf P}^{*}$ denotes the FLR-corrected CGL pressure tensor 
\begin{equation}
{\sf P}^{*} \;\equiv\; {\sf P}_{CGL} \;+\; e\,\Omega\,p_{\bot}\;({\bf I} - \bhat\bhat) \;=\; p_{\|}\,\bhat\bhat \;+\; 
p_{\bot}\,(1 + e\,\Omega)\;({\bf I} - \bhat\bhat), 
\label{eq:CGL_mod}
\end{equation}
the Hermitian conjugate operator $\|\cdots\|^{\dag}$ is defined in Eq.~(\ref{eq:dagger}), and we introduced the scalar field
\begin{equation}
\Lambda \;\equiv\; \x\bdot\pd{\L}{{\bf u}}, \label{eq:Lambda}
\end{equation}
and the vector field
\begin{eqnarray}
\vb{\Gamma} & \equiv & {\bf u}\;\Lambda \;+\; \left(\; {\sf P}^{*} \;-\; 
n\;\pd{{\cal L}}{n}  \;{\bf I} \;\right)\bdot\x \;+\; \delta\phi\;\pd{\L}{(\nabla\phi)} \nonumber \\
 &  &\mbox{}+\; \left[\; \left( \delta\phi\;\nabla_{\bot}\pd{(en\|\phi\|)}{(\nabla_{\bot}^{2}\phi)} \;-\; \nabla_{\bot}\delta\phi\;
\pd{(en\|\phi\|)}{(\nabla_{\bot}^{2}\phi)} \right) \;+\; \cdots \;\right]. \label{eq:Gamma}
\end{eqnarray}
Note that, while the last two terms $\partial_{t}\Lambda + \nabla\bdot\vb{\Gamma}$ in Eq.~(\ref{eq:Eul_Lag}) do not play a role in the variational principle $\int \delta\L\,d^{3}x\,dt = 0$, they play a crucial role in the derivation of exact conservation laws based on the Noether method.

\subsection{Gyrofluid Equations}

\subsubsection{Gyrofluid velocity}

From the Eulerian variation of the Lagrangian density (\ref{eq:Eul_Lag}), the stationarity of the action functional with respect to a arbitrary virtual fluid displacement $\x$ yields the Euler-Poincar\'{e} equation
\begin{eqnarray}
0 & = & \pd{}{t} \pd{\L}{{\bf u}} \;+\; \nabla\bdot\left({\bf u}\;\pd{\L}{{\bf u}} \right) 
\;+\; \nabla{\bf u}\bdot\pd{\L}{{\bf u}} \;+\; \nabla\bdot{\sf P}^{*} \nonumber \\
&  & \;-\; n \left[ \nabla \left( \pd{{\cal L}}{n} \;+\; {1\over 2}T_{\parallel} \;+\; T_{\perp}  \right) \;+\; 
e\,\Omega\;\nabla T_{\bot} \right],
\label{eq:EP}
\end{eqnarray}
which describes the evolution of the gyrofluid velocity ${\bf u}$.  Upon substituting the Lagrangian derivatives 
(\ref{eq:Lag_deriv}) into the Euler-Poincar\'{e} equation (\ref{eq:EP}), and using the fact that the background magnetic field ${\bf B}$ is assumed to be a time-independent nonuniform vector field, we obtain
\begin{eqnarray}
0 & = & mn \left( \pd{u_{\|}}{t} \;+\; {\bf u}\bdot\nabla u_{\|}\right) \bhat \;-\; \frac{en}{c}\;{\bf u}\btimes
{\bf B}^{*} \;+\; en\;\nabla \psi \nonumber \\
&  & \;+\; \nabla p_{\perp} \;+\; T_{\perp}\, \nabla \left(e\,n\;\Omega \right) \;+\; \nabla \left[\left(p_{\Delta} \;-\; e\,\Omega \;p_{\perp} \right)\bhat\bhat \right],
\label{eq:gyro_velocity}
\end{eqnarray}

where $p_{\Delta}=p_{\parallel}-p_{\perp}$, and we have introduced the following definitions
\begin{equation}
\left. \begin{array}{rcl}
{\bf B}^{*} & \equiv & {\bf B} \;+\; (mc/e)\;u_{\|}\;\nabla\btimes\bhat \\
 &  & \\
B_{\|}^{*} & \equiv & {\bf B}^{*}\bdot\bhat \;=\; B \;+\; (mc/e)\;u_{\|}\;\bhat\bdot\nabla\btimes\bhat \\
&  & \\
\bhat^{*} & \equiv & {\bf B}^{*}/B_{\|}^{*} \;=\; \bhat \;+\; mu_{\|}\;{\bf C}
\end{array} \right\},
\label{eq:gyro_defs}
\end{equation}
with the magnetic-curvature term ${\bf C}$ defined as ${\bf C} \equiv (c/eB_{\|}^{*})\,\bhat\btimes(\bhat\bdot\nabla
\bhat)$.

Eq.~(\ref{eq:gyro_velocity}) can also be written in a more compact form by introducing the gradient of $\psi$ evaluated at constant perpendicular temperature,
$\nabla^{\top}\psi$, defined through  Eqs.~(\ref{eq:phi_gyro}) and (\ref{eq:Omega_def}) as
\[ \nabla\|\phi\| \;=\; \|\nabla\phi\| \;+\; \pd{\|\phi\|}{B}\;\nabla B \;+\; \pd{\|\phi\|}{T_{\bot}}\;\nabla T_{\bot} 
\;\equiv\; \nabla^{\top}\|\phi\| \;+\; \Omega\;\nabla T_{\bot}, \]
where $\partial\|\phi\|/\partial B \equiv -\,2\,(T_{\bot}/B)\,\Omega$. Then Eq.~(\ref{eq:gyro_velocity}) can be written as:
\begin{equation}
0 \;=\; mn \left( \pd{u_{\|}}{t} \;+\; {\bf u}\bdot\nabla u_{\|}\right) \bhat \;-\; \frac{en}{c}\;{\bf u}\btimes
{\bf B}^{*} \;+\; en\;\nabla^{\top}\psi \;+\; \nabla\bdot{\sf P}^{*}.
\end{equation}

Note that the vector field ${\bf B}^{*}$ is NOT a divergenceless field (since $\nabla u_{\|} \neq 0$), as is common to all guiding-center and gyrocenter Hamiltonian models (in which parallel gyrofluid velocity $u_{\|}$ is replaced with the parallel guiding-center velocity $v_{\|}$). 

It is easy to see that Eq.~(\ref{eq:gyro_velocity}) can be divided into two equations: one equation that expresses the gyrofluid velocity ${\bf u}$ in terms of the gyrofluid moments $(n,u_{\|},p_{\|},p_{\bot})$ and the electrostatic potential $\phi$, and one equation that describes the time evolution of the parallel gyrofluid velocity $u_{\|}$. The first equation is obtained by taking the cross-product of Eq.~(\ref{eq:gyro_velocity}) with $\bhat$, which
yields the following expression for the gyrofluid velocity
\begin{equation}
{\bf u} \;\equiv\; u_{\|}\,\bhat^{*} \;+\; \left. \left. \frac{c\bhat}{en\,B_{\|}^{*}}\btimes\right\{ e\,n\,\nabla \psi\;+\; \nabla p_{\perp} \;+\; T_{\perp}\, \nabla \left(e\,n\;\Omega \right) \;+\; \nabla \left[\left(p_{\Delta} \;-\; e\,\Omega \;p_{\perp} \right)\bhat\bhat \right]\right\},
\label{eq:u_vector}
\end{equation}
Therefore, according to Eq.~\ref{eq:u_vector}, the gyrofluid velocity ${\bf u}$, consists of the parallel velocity $u_{\|} \equiv {\bf u}\bdot\bhat$ and the following perpendicular gyrofluid velocities
\begin{equation}
\left. \begin{array}{rcl}
{\bf u}_{D} & \equiv & (c/en\,B_{\|}^{*})\,\bhat\btimes\nabla p_{\bot} \\
 &  & \\
{\bf u}_{\psi} & \equiv & (c/B_{\|}^{*})\,\bhat\btimes\nabla\psi \\
 &  & \\
{\bf u}_{C} & \equiv & (p_{\Delta}/n)\;{\bf C} 
\end{array} \right\},
\label{eq:gyro_perp}
\end{equation}
which represent the diamagnetic velocity, the generalized $E\times B$ velocity, and the curvature-drift velocity, respectively, and the following gyro-potential FLR corrections 
\begin{equation} 
{\bf w}_{\Omega} \;\equiv\; \frac{cT_{\bot}\bhat}{nB_{\|}^{*}}\btimes\nabla(n\Omega) \;\;\;{\rm and}\;\;\; 
{\bf w}_{C} \;\equiv -\;e\Omega\,T_{\bot}\;{\bf C}
\label{eq:gyro_FLR}
\end{equation}
to the $E\times B$ velocity and curvature-drift velocity, respectively.

\subsubsection{Evolution equation for $u_{\|}$}

The evolution equation for parallel gyrofluid velocity $u_{\|}$ can be derived by taking the dot-product of 
Eq.~(\ref{eq:gyro_velocity}) with $\bhat^{*}$:
\begin{eqnarray}
mn\,\left( \pd{u_{\|}}{t} \;+\; {\bf u}\bdot\nabla u_{\|} \right) & = & -\; \left. \left. \bhat^{*}\bdot\right[\; en\;\nabla\psi \;+\; \nabla p_{\bot} \;+\; T_{\bot}\;\nabla(en\Omega) \;\right] \nonumber \\
 &  &\mbox{}-\;\nabla\bdot\left[\;(p_{\Delta} - e\Omega\,p_{\bot})\;\bhat\;\right],
\label{eq:u_parallel}
\end{eqnarray}
where we have used the identity (valid for any function $f$)
\[ \nabla\bdot(f\;\bhat\bhat) \;=\; \left[\;\nabla\bdot(f\;\bhat)\;\right] \bhat \;+\; f\;(\bhat\bdot\nabla\bhat), \]
so that we find $\bhat^{*}\bdot\nabla\bdot(f\;\bhat\bhat) = \nabla\bdot(f\;\bhat)$. The gyrofluid equation 
(\ref{eq:u_parallel}) for $u_{\|}$ includes terms associated with the parallel electric field and its FLR corrections as well as parallel thermal forces.

The set of gyrofluid equations of motion for the four gyrofluid moments $(n,p_{\|},p_{\bot},u_{\|})$ are, thus, given by Eqs.~(\ref{eq:continuity}), (\ref{eq:p_parq})-(\ref{eq:p_perpq}), and (\ref{eq:u_parallel}), respectively. Each of these gyrofluid equations involves the advection operator ${\bf u}\bdot\nabla$ and the divergence $\nabla\bdot{\bf u}$; it is the diamagnetic part of the advection operator that must be eliminated from Eqs.~(\ref{eq:p_parq})-(\ref{eq:p_perpq}) and (\ref{eq:u_parallel}) by adding suitable diamagnetic fluxes.

\subsection{Polarization Equation}

The polarization equation for $\phi$ can be found by considering the terms in the variation (\ref{eq:Eul_Lag}) that involve the variation of the potential $\delta \phi$. Here, we must remember to add up the contributions of all the species to the total Lagrangian density. 

Using the variation of the Lagrangian density (\ref{eq:Eul_Lag}), we obtain the Euler-Lagrange equation for $\phi$:
\begin{equation}
0 \;=\; \sum_{j}\;\|(en)_{j}\|^{\dag} \;+\; \nabla\bdot\pd{\L}{(\nabla\phi)},
\label{eq:EL_phi}
\end{equation}
where summation over particle species $(\sum_{j})$ is shown explicitly. Using Eq.~(\ref{eq:Lag_total}), we find
\begin{equation}
\pd{\L}{(\nabla\phi)} \;=\; \frac{1}{4\pi}\;\nabla\phi \;+\;\sum_{j}\; \frac{(nm)_{j}c^{2}}{B^{2}}\;\nabla_{\bot}\phi, 
\label{eq:grad_phi}
\end{equation}
so that the Euler-Lagrange (\ref{eq:EL_phi}) becomes the polarization equation
\begin{equation}
\sum_{j}\left[\; \|(en)_{j}\|^{\dag} \;+\; \nabla_{\bot}\bdot \left( \frac{(nm)_{j}c^2}{B^{2}}\;\nabla_{\bot}\phi\right) \right] \;+\; \frac{1}{4\pi}\,\nabla^2\phi \;=\; 0.
\label{eq:polarization}
\end{equation} 
Comparing Eq.~(\ref{eq:polarization}) with our previous one-temperature model \cite{ours}, it is clear that the polarization equation is not affected by the introduction of the parallel temperature in the model. This arises from the fact that the terms that contain the gyrofluid electrostatic potential (\ref{eq:phi_gyro}) do not depend on 
$p_{\parallel}$. Further details concerning Eq.~(\ref{eq:polarization}) can be found in Ref.~\cite{ours}.

Lastly, we note that two versions for the function ${\cal P}(x)$, appearing in the definition of the gyrofluid scalar potential (\ref{eq:phi_gyro}), are commonly used in gyrofluid applications: the function ${\cal P}(x) = e^{x}$ or its 
Pad\'{e} approximant ${\cal P}(x) = (1 - x)^{-1}$. It is important to note that, whatever form is adopted for the gyrofluid scalar potential $\|\phi\|$, it must be used consistently throughout the model in order to ensure energy conservation.

\section{Energy conservation law}

In this Section. we present the local and global forms of the energy conservation law, as they arise from the application of the Noether method. For this purpose, we point out that, as a result of the variational principle $\int \delta\L\,d^{3}x\,dt = 0$, the only remaining terms in the Eulerian variation of the Lagrangian density (\ref{eq:Eul_Lag}) become the Noether equation
\begin{equation}
\delta\L \;=\; \pd{\Lambda}{t} \;+\; \nabla\bdot\vb{\Gamma},
\label{eq:Noether}
\end{equation}
where $\Lambda$ and $\vb{\Gamma}$ are defined in Eqs.~(\ref{eq:Lambda})-(\ref{eq:Gamma}). The energy and momentum conservation laws are derived from the Noether equation (\ref{eq:Noether}) by considering infinitesimal time and space translations, respectively. In the present work, we focus our attention on the local and global energy conservation laws associated with our electrostatic gyrofluid model.

\subsection{Local energy conservation law}

We derive the local form of the energy conservation law from the Noether equation (\ref{eq:Noether}) by considering infinitesimal time translations
$t \rightarrow t + \delta t$, from which we obtain the following expressions for the virtual fluid displacement $\x$ and the Eulerian variations
$\delta\phi$ and $\delta\L$:
\begin{equation}
\x \;=\; -\,{\bf u}\,\delta t, \;\;\; \delta\phi \;=\; -\,\delta t\;\partial_{t}\phi,\;\;\;{\rm and}\;\;\; \delta\L \;=\; -\,\delta t\;\partial_{t}\L.
\label{eq:time}
\end{equation}
Inserting these expressions into Eqs.~(\ref{eq:Lambda})-(\ref{eq:Gamma}), we obtain
\begin{eqnarray}
\Lambda & \equiv & -\;\delta t\;\left( {\bf u}\bdot\pd{\L}{{\bf u}} \right), \label{eq:Lambda_time} \\
\vb{\Gamma} & \equiv & -\;\delta t \left[\; {\bf u} \left( {\bf u}\bdot\pd{\L}{{\bf u}} \;-\; n\;\pd{{\cal L}}{n} \right) \;+\; {\sf P}^{*}\bdot{\bf u} \;+\; \pd{\phi}{t}\;
\pd{\L}{(\nabla\phi)} \right. \nonumber \\
 &  &\left.+\; \left( \pd{\phi}{t}\;\nabla_{\bot}\pd{(en\|\phi\|)}{(\nabla_{\bot}^{2}\phi)} \;-\; \nabla_{\bot}
\pd{\phi}{t}\;\pd{(en\|\phi\|)}{(\nabla_{\bot}^{2}\phi)} \;+\; \cdots \right) \;\right], 
\label{eq:Gamma_time}
\end{eqnarray}
where summation over fluid species is implied wherever appropriate. By combining these expressions, we arrive at the {\it primitive} form of the local energy conservation law;
\[ \pd{\varepsilon^{\prime}}{t} \;+\; \nabla\bdot{\bf S}^{\prime} \;=\; 0, \]
where the primitive energy density is
\begin{equation}
\varepsilon^{\prime} \;\equiv\; {\bf u}\bdot\pd{\L}{{\bf u}} \;-\; \L \;=\; \frac{1}{2}\,nm\,u_{\|}^{2} \;+\; p_{\bot} \;+\; \frac{p_{\|}}{2} \;+\; 
\left( en\,\psi \;-\; \frac{|{\bf E}|^{2}}{8\pi} \right), \label{eq:E_prim}
\end{equation}
and the primitive energy-density flux is
\begin{eqnarray}
{\bf S}^{\prime} & \equiv & {\bf u} \left( {\bf u}\bdot\pd{\L}{{\bf u}}
\;-\; n\;\pd{{\cal L}}{n} \right) \;+\; {\sf P}^{*}\bdot{\bf u} \;+\; 
\pd{\phi}{t}\;\pd{\L}{(\nabla\phi)} \nonumber \\
 &  &\mbox{}+\; \left( \pd{\phi}{t}\;\nabla_{\bot}\pd{(en\|\phi\|)}{(\nabla_{\bot}^{2}\phi)} \;-\;
\nabla_{\bot}\pd{\phi}{t}\;\pd{(en\|\phi\|)}{(\nabla_{\bot}^{2}\phi)} \;+\; \cdots \right) \nonumber \\
& = & {\bf u} \left( \frac{1}{2}\,nm\,u_{\|}^{2} \;+\; p_{\bot} \;+\; \frac{p_{\|}}{2} \;+\; en\,\psi \right) \;+\; 
{\sf P}^{*}\bdot{\bf u} \;+\; \pd{\phi}{t}\pd{\L}{(\nabla\phi)} \nonumber \\
 &  &\mbox{}+\; \left[\; \pd{\phi}{t}\;\nabla_{\bot}\left( \pd{(en\;\|\phi\|)}{(\nabla_{\bot}^{2}\phi)} \;+\; \cdots \right) \;-\; 
\nabla_{\bot}\pd{\phi}{t}\;\left( \pd{(en\;\|\phi\|)}{(\nabla_{\bot}^{2}\phi)} \;+\; \cdots \right) \;\right]. 
\label{eq:S_prim}
\end{eqnarray}
In order to arrive at the final form of the energy conservation law, we need to rearrange terms in Eq.~(\ref{eq:E_prim}). By substituting the polarization equation (\ref{eq:EL_phi}) into Eq.~(\ref{eq:dag_identity}), we find
\begin{equation}
en\;\|\phi\| \;=\; \frac{|{\bf E}|^{2}}{4\pi} \;+\; mn\;|{\bf u}_{E}|^{2} \;-\; \nabla\bdot{\bf D},
\label{eq:switch_phi}
\end{equation}
where
\begin{equation}
{\bf D} \;\equiv\; \phi\;\pd{\L}{(\nabla\phi)} \;-\; \pd{(en\;\|\phi\|)}{(\nabla_{\bot}^{2}\phi)}\;\nabla_{\bot}\phi \;+\; 
\phi\;\nabla_{\bot}\left(\pd{(en\;\|\phi\|)}{(\nabla_{\bot}^{2}\phi)}\right) \;+\; \cdots
\label{eq:energy_gauge}
\end{equation}
so that the last terms in Eq.~(\ref{eq:E_prim}) become
\begin{equation} 
en\,\psi \;-\; \frac{|{\bf E}|^{2}}{8\pi} \;=\; \frac{1}{2}\,nm\;|{\bf u}_{E}|^{2} \;+\; \frac{|{\bf E}|^{2}}{8\pi} \;-\; \nabla\bdot{\bf D}.
\label{eq:field_en}
\end{equation}
Hence, we express the primitive energy density (\ref{eq:E_prim}) as $\varepsilon^{\prime} \equiv \varepsilon \;-\; \nabla\bdot{\bf D}$, where the final form of the energy density is defined as
\begin{equation}
\varepsilon \;\equiv\; \frac{1}{2}\,nm\,\left( u_{\|}^{2} \;+\; |{\bf u}_{E}|^{2} \right) \;+\; p_{\bot} \;+\; \frac{p_{\|}}{2} \;+\; 
\frac{|{\bf E}|^{2}}{8\pi},
\label{eq:energy_density}
\end{equation}
and we obtain the local form of the energy conservation law
\begin{equation}
\pd{\varepsilon}{t} \;+\; \nabla\bdot{\bf S} \;=\; 0,
\label{eq:ECL}
\end{equation}
where the final form of the energy density flux is defined as
\begin{equation}
{\bf S} \;\equiv\; {\bf S}^{\prime} \;-\; \pd{{\bf D}}{t}.
\label{eq:energy_flux}
\end{equation}
After some partial cancelations, the final form of the energy density flux is
\begin{equation}
{\bf S} \;=\; {\bf u} \left( \frac{1}{2}\,nm\,u_{\|}^{2} \;+\; p_{\bot} \;+\; \frac{p_{\|}}{2} \;+\; en\,\psi \right) 
\;+\; {\sf P}^{*}\bdot{\bf u} \;+\; {\bf S}_{\phi},
\label{eq:S_final}
\end{equation}
where ${\sf P}^{*}\bdot{\bf u} = p_{\bot}\,(1 + e\Omega)\,{\bf u} + (p_{\Delta} - e\Omega\,p_{\bot})\;u_{\|}\,\bhat$ and the electrostatic energy density flux is defined as
\begin{equation} 
{\bf S}_{\phi} \;=\; -\; \phi\;\pd{}{t}\left( \pd{\L}{(\nabla\phi)} \;+\; \nabla_{\bot}
\pd{(en\;\|\phi\|)}{(\nabla_{\bot}^{2}\phi)} \;+\; \cdots \right) \;+\; \nabla_{\bot}\phi\;\pd{}{t}\left( 
\pd{(en\;\|\phi\|)}{(\nabla_{\bot}^{2}\phi)} \;+\; \cdots \right).
\label{eq:electric_flux}
\end{equation}
In the next Section, after heat fluxes are inserted back into the pressure evolution equations and diamagnetic cancellations are performed with the addition of terms in the gyrofluid equations, the local energy conservation law (\ref{eq:ECL}) is converted into a new energy equation $\partial_{t}\varepsilon + \nabla\bdot{\bf S}^{*} = 0$, in which heat fluxes and diamagnetic-cancellation terms result in a modified energy density flux ${\bf S}^{*}$. This new form ensures that the total energy ${\cal E} = \int \varepsilon\,d^{3}x$ satisfies the global energy conservation law $d{\cal E}/dt 
= 0$.

\subsection{Global Energy Conservation Law}

The global energy conservation law $d{\cal E}/dt = 0$ can be derived from the local energy conservation law 
(\ref{eq:ECL}) by integrating it over space. Here, the global energy is defined as
\begin{equation}
{\cal E} \;=\; \int d^{3}x \left[\; \frac{1}{2}\,nm\,u_{\|}^{2} \;+\; \left( p_{\bot} \;+\; \frac{p_{\|}}{2} \right) 
\;+\; \left( \frac{1}{2}\,nm|{\bf u}_{E}|^{2} \;+\; \frac{|{\bf E}|^{2}}{8\pi} \right) \;\right].
\label{eq:energy_global}
\end{equation}
In this form, the parallel kinetic energy, the internal energy, and the electric field energy explicitly appear. In a later section, we will present the time evolution of each of the separate terms that constitute the energy conservation law in order to identify the energy-exchange processes that allow the transfer of energy between the three types of gyrofluid (parallel kinetic, internal, and field) energies. 

\section{Diamagnetic cancelations and Energy conservation}

The gyrofluid velocity (\ref{eq:gyro_velocity}) contains the diamagnetic
velocity ${\bf u}_{D}$. Since the gyrofluid moment-equations are derived
by inserting the gyrofluid velocity ${\bf u}$ in the Lagrangian
constraints, diamagnetic advection terms appear in the equations of
evolution for the parallel velocity and the two anisotropic
pressures. More specifically, the momentum equation for $u_{\|}$
contains the term ${\bf u}_{D}\bdot \nabla u_{\|}$, and the parallel and
perpendicular pressure equations contain a combination of the terms
${\bf u}_{D}\bdot \nabla p$ and $p\nabla \bdot {\bf u}_{D}$ (where here
p is either $p_{\parallel}$ or $p_{\perp}$). These diamagnetic-advection
terms should be canceled in the gyrofluid evolution equations by the
introduction of appropriate terms containing higher-order moments
\cite{ours}.  These correspond to the FLR corrections to perpendicular
fluxes in conventional fluid models \cite{grad,brag}, but arise
naturally due to grad-B and curvature drifts in the moment-based
derivation of local gyrofluid models \cite{beer}.

Since the higher-order moment terms cannot be derived from the Lagrangian, the diamagnetic cancelations must be done manually. However, there exists a constraint in the addition of higher-order moment terms, namely, that the global energy conservation law should not be altered. Here, the terms added are derived from the Vlasov equation, and FLR corrections are then introduced to conserve energy. Thus, the final moment equations are not directly derived from the Lagrangian, but they still conserve energy exactly. 

\subsection{Parallel Gyrofluid Dynamics}

The diamagnetic cancelation needed for the parallel momentum equation (\ref{eq:u_parallel}) involves the addition of the term $-\,\nabla\bdot
\vb{\Pi}_{\|}^{*}$, associated with the non-diagonal part of the pressure tensor \cite{hinton}, on the right side of Eq.~(\ref{eq:u_parallel}). From Vlasov theory, the diamagnetic-cancelation term is found to be \cite{belova}:
\begin{equation}
-\;\nabla \bdot \vb{\Pi}_{\|}^{*} \;=\; -\;\nabla \bdot \left( p_{\perp}\; \frac{mc\bhat}{eB_{\|}^{*}}\btimes 
\nabla u_{\|} \right) \;=\; mn{\bf u}_{D}\bdot\nabla u_{\|} \;+\; p_{\bot}\;\K(mu_{\|}),
\label{eq:diacan_upar}
\end{equation}
where the {\it magnetic} differential operator $\K(\cdots)$ is defined by the identity
\begin{equation} 
\nabla\bdot\left( g\;\frac{c\bhat}{eB_{\|}^{*}}\btimes\nabla f \right) \;=\; \frac{c\bhat}{eB_{\|}^{*}}\bdot\nabla f
\btimes\nabla g \;-\; g\,\K(f),
\label{eq:K_id}
\end{equation}
valid for arbitrary functions $f$ and $g$. As written here, Eq.~(\ref{eq:diacan_upar}) conserves energy by itself, since 
$u_{\|}\;\nabla\bdot\vb{\Pi}_{\|}^{*} = \nabla\bdot(u_{\|}\;\vb{\Pi}_{\|}^{*})$. Here, we should note that the substitution of $B$ to $\bpar*$ is done here, and throughout this Section, to make the diamagnetic cancelation exact.

The time evolution of the parallel kinetic energy density is, therefore, expressed as
\begin{eqnarray}
\pd{}{t} \left( \frac{mn}{2}\;u_{\|}^{2} \right) & = & -\;\nabla\bdot \left( \frac{mn}{2}\;u_{\|}^{2}\,{\bf u} \;+\; 
u_{\|}\;\vb{\Pi}_{\|}^{*} \right) \nonumber \\
&  & \;-\; {\bf u}\bdot\left\{ en\nabla \psi \;+\; \nabla p_{\perp}\;+\; T_{\perp}\nabla (e\,n\,\Omega)+\nabla \left[\left(\;p_{\Delta}-e\,\Omega\,p_{\perp}\;\right)\bhat\bhat\right]\right\},
\label{eq:par_kinetic}
\end{eqnarray}
where we used the expression (\ref{eq:u_vector}) for the gyrofluid velocity to write $u_{\|}\,\bhat^{*}\bdot\{\cdots\}
= {\bf u}\bdot\{\cdots\}$ for the second term on the right side of Eq.~(\ref{eq:par_kinetic}).

\subsection{Internal Energy}

To consider the time evolution of the internal energy $p_{\bot} + p_{\|}/2$, we rewrite the two pressure equations 
(\ref{eq:p_parq})-(\ref{eq:p_perpq}) in the form
\begin{eqnarray}
\pd{p_{\|}}{t} \;+\; \nabla\bdot( p_{\|}\;{\bf u}) \;+\; 2\,p_{\parallel}\;\bhat\bhat:\nabla{\bf u} & = & -\;2\,
\nabla\bdot{\bf q}_{\parallel\perp} \;+\; 4\,{\bf q}_{\parallel \perp}\bdot (\bhat\bdot \nabla\bhat) \;+\; 2\,Q_{\|}, 
\label{eq:ppar_dc} \\
\pd{p_{\bot}}{t} \;+\; \nabla\bdot(p_{\bot}\;{\bf u}) \;+\; p_{\perp}\;({\bf I} - \bhat\bhat):\nabla{\bf u} & = & 
-\;\nabla\bdot{\bf q}_{\perp\perp} \;-\; 2\,{\bf q}_{\parallel \perp}\bdot (\bhat\bdot \nabla\bhat) \;+\; Q_{\bot}, 
\label{eq:pper_dc}
\end{eqnarray}
where $Q_{\|}$ and $Q_{\bot}$ are additional terms (to be determined later) associated with FLR corrections to the electrostatic scalar field $\phi$. 

To zeroth order in the electrostatic potential $\phi$, the heat fluxes are derived from the Vlasov equation directly and are found in Ref.~\cite{weiland} to be expressed as
\begin{equation}
{\bf q}_{\parallel \perp} \;=\; \frac{1}{2}\;\frac{cp_{\perp}\bhat}{eB_{\|}^{*}} \btimes \nabla T_{\parallel} \;+\; p_{\|}\;{\bf u}_{C},
\label{eq:qpar}
\end{equation}
and
\begin{equation}
{\bf q}_{\perp \perp} \;=\; 2\;\frac{cp_{\perp}\bhat}{eB_{\|}^{*}}\btimes \nabla T_{\perp}.
\label{eq:qperp}
\end{equation}
Inserting these diamagnetic heat fluxes, using the definition (\ref{eq:K_id}), we find
\begin{eqnarray}
-\,2\,\nabla \bdot {\bf q}_{\parallel \perp} \;+\; 4\,{\bf q}_{\parallel \perp}\bdot(\bhat\bdot\nabla\bhat) & = &
n{\bf u}_{D}\bdot\nabla T_{\|} \;+\; p_{\perp}\;\K(T_{\parallel}) \nonumber \\
 &  &\mbox{}-\; 2\,p_{\bot}\;\c \bdot \nabla T_{\parallel} \;-\; 2\;\nabla \bdot \left( T_{\|}\,p_{\Delta}\;\c \right)
\label{eq:addpar}
\end{eqnarray}
for the parallel pressure equation (\ref{eq:p_parq}), and
\begin{equation}
-\;\nabla \bdot {\bf q}_{\perp \perp} \;-\; 2\,{\bf q}_{\parallel \perp}\bdot(\bhat\bdot\nabla\bhat) \;=\; 
2\;n{\bf u}_{D}\bdot\nabla T_{\perp} \;+\; 2\;p_{\perp}\;\K(T_{\perp}) \;+\; p_{\perp}\;\c \bdot\nabla T_{\parallel}
\label{eq:addperp}
\end{equation}
for the perpendicular pressure equation (\ref{eq:p_perpq}). 

Using Eqs.~(\ref{eq:ppar_dc}) and (\ref{eq:pper_dc}), the time evolution of the internal energy can, therefore, be expressed as
\begin{eqnarray}
\pd{}{t} \left( \frac{p_{\|}}{2} \;+\; p_{\bot} \right) & = & -\;\nabla\bdot\left[\; \left( \frac{p_{\|}}{2} \;+\; 
p_{\bot} \right) {\bf u} \;+\; {\sf P}_{CGL}\bdot{\bf u} \;+\; \left( {\bf q}_{\|\bot} + {\bf q}_{\bot\bot} \right) \;\right] \nonumber \\
 &  &\mbox{}+\; {\bf u}\bdot\nabla\bdot{\sf P}_{CGL} \;+\; \left( Q_{\|} \;+\; Q_{\bot} \right),
\label{eq:internal}
\end{eqnarray}
where suitable energy-conserving expressions for the FLR-corrected heat fluxes $Q_{\|}$ and $Q_{\bot}$ are now determined by considering the expression for the time evolution of the electrostatic field energy.

\subsection{Electrostatic Field Energy}

By making use of the electrostatic field energy equation (\ref{eq:field_en}), we write the following expression for the time evolution of the electrostatic field energy
\begin{eqnarray}
\pd{}{t} \left( \frac{mn}{2}\;|{\bf u}_{E}|^{2} \;+\; \frac{|{\bf E}|^{2}}{8\pi} \right) & = & \nabla\bdot\pd{{\bf D}}{t} \;+\; \pd{}{t} \left( en\,\psi \;-\; \frac{|{\bf E}|^{2}}{8\pi} \right) \nonumber \\
 & = & e\,(\psi \;-\; T_{\bot}\,\Omega)\;\pd{n}{t} \;+\; e\Omega\;\pd{p_{\bot}}{t} \;-\; \nabla\bdot{\bf S}_{\phi},
\label{eq:field_evol}
\end{eqnarray}
where we made use of the polarization equation (\ref{eq:polarization}) and the definition (\ref{eq:electric_flux}) for 
${\bf S}_{\phi}$. Here, using the gyrofluid continuity (\ref{eq:continuity}), we obtain
\[ e\,(\psi \;-\; T_{\bot}\Omega)\;\pd{n}{t} \;=\; -\;\nabla\bdot\left[\; e\,(n\psi \;-\; p_{\bot}\,\Omega)\;{\bf u} 
\;\right] \;+\; en\,{\bf u}\bdot\left[\; \nabla\psi \;-\; \nabla(T_{\bot}\Omega)\;\right], \]
and, using the perpendicular pressure equation (\ref{eq:pper_dc}), we obtain
\begin{eqnarray*} 
e\Omega\;\pd{p_{\bot}}{t} & = & -\;e\Omega\;\nabla\bdot(p_{\bot}\;{\bf u}) \;-\; {\sf P}_{\Omega}:\nabla{\bf u} \;-\; e\Omega\;\left[\; 
\nabla\bdot{\bf q}_{\perp\perp} \;+\; 2\,{\bf q}_{\parallel \perp}\bdot (\bhat\bdot \nabla\bhat) \;-\; Q_{\bot} \;\right] \nonumber \\
 & = & -\;\nabla\bdot\left(\; e\Omega\,p_{\bot}\;{\bf u} \;+\; {\sf P}_{\Omega}\bdot{\bf u} \;\right) \;+\; {\bf u}\bdot\left[\; p_{\bot}\;
\nabla(e\Omega) \;+\; \nabla\bdot{\sf P}_{\Omega} \;\right] \\
 &  &\mbox{}-\; e\Omega\;\left[\; \nabla\bdot{\bf q}_{\perp\perp} \;+\; 2\,{\bf q}_{\parallel \perp}\bdot (\bhat\bdot \nabla\bhat) \;-\; Q_{\bot}
\;\right], 
\end{eqnarray*}
where ${\sf P}_{\Omega} \equiv {\sf P}^{*} - {\sf P}_{CGL} = e\Omega\,p_{\bot}\;({\bf I} - \bhat\bhat)$ denotes the FLR-correction to the CGL pressure tensor. Hence, the first two terms on the right side of Eq.~(\ref{eq:field_evol}) can be written as
\begin{eqnarray}
e\,(\psi \;-\; T_{\bot}\,\Omega)\;\pd{n}{t} \;+\; e\Omega\;\pd{p_{\bot}}{t} & = & -\;\nabla\bdot\left(\; e\,n\psi\,{\bf u} \;+\; {\sf P}_{\Omega}\bdot
{\bf u} \;\right) \nonumber \\
&  & \;+\; {\bf u}\bdot\left( en\;\nabla \psi \;-\; en\,\Omega \nabla T_{\perp} \;+\; \nabla\bdot{\sf P}_{\Omega} \right) \nonumber \\
 &  &\mbox{}-\; e\Omega\;\left[\; \nabla\bdot{\bf q}_{\perp\perp} \;+\; 2\,{\bf q}_{\parallel \perp}\bdot (\bhat\bdot \nabla\bhat) \;-\; Q_{\bot}
\;\right].
\label{eq:field_Omega}
\end{eqnarray}
We now require that, in order for the last term $Q_{\Omega} \equiv Q_{\|} + Q_{\bot}$ appearing on the right side of Eq.~(\ref{eq:internal}) to cancel the last three terms on the right side of Eq.~(\ref{eq:field_Omega}), the latter terms must be written up to an exact spatial divergence as
\[ -\; e\Omega\;\nabla\bdot{\bf q}_{\perp\perp} \;-\; 2\,e\Omega\;{\bf q}_{\parallel \perp}\bdot (\bhat\bdot \nabla\bhat) \;+\; e\Omega\;Q_{\bot} \;=\; 
-\;Q_{\Omega} \;-\; \nabla\bdot{\bf S}_{\Omega}. \]
Hence, we set the additional terms $Q_{\|}$ and $Q_{\bot}$ in Eqs.~(\ref{eq:internal}) and (\ref{eq:field_Omega}) to be
\begin{equation}
\left. \begin{array}{rcl}
Q_{\|} & = & 2\;e\Omega\;{\bf q}_{\|\bot}\bdot\left(\bhat\bdot\nabla\bhat\right) \\
 &  & \\
Q_{\bot} & = & -\;{\bf q}_{\bot\bot}\bdot\nabla(e\Omega) \;-\; \nabla\bdot\left( {\bf q}_{\bot\bot}\;e\Omega\right)
\end{array} \right\},
\label{eq:Q_parperp}
\end{equation}
so that 
\begin{equation}
Q_{\Omega} \;\equiv\; Q_{\|} \;+\; Q_{\bot} \;=\; n{\bf w}_{C}\bdot\nabla T_{\|} \;+\; 2\,e\Omega\,p_{\bot}\,\K(T_{\bot}) \;+\; 2n\,
\left( 2\,{\bf w}_{\Omega} - e\Omega\,{\bf u}_{D}\right)\bdot\nabla T_{\bot},
\label{eq:Q_Omega}
\end{equation}
and ${\bf S}_{\Omega} \equiv e\Omega\,(2 + e\Omega)\;{\bf q}_{\bot\bot}$, with ${\bf q}_{\bot\bot}$ defined in Eq.~(\ref{eq:qperp}). Combining Eqs.~(\ref{eq:field_Omega})-(\ref{eq:Q_Omega}) into Eq.~(\ref{eq:field_evol}), the time evolution of the electrostatic field energy is, therefore, expressed as
\begin{eqnarray}
\pd{}{t} \left( \frac{mn}{2}\;|{\bf u}_{E}|^{2} \;+\; \frac{|{\bf E}|^{2}}{8\pi} \right) & = & -\;\nabla\bdot\left(\; 
e\,n\psi\,{\bf u} \;+\; {\sf P}_{\Omega}\bdot{\bf u} \;+\; {\bf S}_{\phi} \;+\; {\bf S}_{\Omega} \;\right) \nonumber \\
 &  &\mbox{}+\; {\bf u}\bdot\left( en\;\nabla \psi \;-\; en\,\Omega \nabla T_{\perp} \;+\; \nabla\bdot{\sf P}_{\Omega} \right) \;-\; Q_{\Omega}.
\label{eq:field_energy}
\end{eqnarray}

\subsection{Explicit form of the energy conservation law}

When diamagnetic cancellations and heat fluxes are introduced into the four-moment gyrofluid equations, the local energy conservation law (\ref{eq:ECL}) is modified. By combining the evolution equations for the parallel kinetic energy 
(\ref{eq:par_kinetic}), the internal energy (\ref{eq:internal}), and the electrostatic field energy 
(\ref{eq:field_energy}), the local energy conservation law (\ref{eq:ECL}) becomes the local energy equation
\begin{equation}
\pd{\varepsilon}{t} \;+\; \nabla\bdot{\bf S} \;=\; -\;\nabla\bdot\left[\; u_{\|}\;\vb{\Pi}_{\|}^{*} \;+\; {\bf q}_{\|\bot}
\;+\; (1 + e\Omega)^{2}\;{\bf q}_{\bot\bot} \;\right],
\label{eq:energy_heat}
\end{equation}
which ensures that the total energy ${\cal E} = \int \varepsilon d^{3}x$ still satisfies the global energy conservation law $d{\cal E}/dt = 0$.

We now identify the energy-exchange processes that transfer energy between the three different types of gyrofluid energy. First, we write down expressions describing the time evolution of each type of energy (e.g., parallel kinetic energy, internal energy, and field energy). Thus, the contribution of each species to the integrated parallel kinetic energy 
(\ref{eq:par_kinetic}) is
\begin{equation}
\frac{d}{dt} \int \left( \frac{mn}{2}\;u_{\|}^{2} \right) d^{3}x \;=\; -\;\int {\bf u}\bdot\left( en\;\nabla \psi \;-\; en\,\Omega \nabla T_{\perp}
\;+\; \nabla\bdot{\sf P}^{*} \right) d^{3}x,
\label{eq:parkin_ex}
\end{equation}
the contribution of each species to the integrated internal energy (\ref{eq:internal}) is
\begin{equation}
\frac{d}{dt} \int \left( \frac{p_{\|}}{2} \;+\; p_{\bot} \right) d^{3}x \;=\; \int \left[\; {\bf u}\bdot\left( 
\nabla\bdot{\sf P}_{CGL}\right) \;+\; Q_{\Omega} \;\right] d^{3}x,
\label{eq:internal_ex}
\end{equation}
and the contribution of each species to the integrated electrostatic field energy (\ref{eq:field_energy}) is
\begin{equation}
\frac{d}{dt} \int \left( \frac{mn}{2}\;|{\bf u}_{E}|^{2} \;+\; \frac{|{\bf E}|^{2}}{8\pi} \right) d^{3}x \;=\; \int
\left[\; {\bf u}\bdot\left( en\;\nabla \psi \;-\; en\,\Omega \nabla T_{\perp} \;+\; \nabla\bdot{\sf P}_{\Omega} \right) \;-\; Q_{\Omega} \;\right] 
d^{3}x.
\label{eq:field_ex}
\end{equation}
The terms on the right side of Eqs.~(\ref{eq:parkin_ex})-(\ref{eq:field_ex}) appear in pairs with opposite sign, and give the energy-exchange processes. For example, the FLR-correction heat flux $Q_{\Omega}$ is involved in energy exchange between the electrostatic field energy and the internal energy, while the FLR-correction pressure tensor ${\sf P}_{\Omega}$ is involved in energy exchange between the electrostatic field energy and the parallel kinetic energy. The contributions from $en\;\nabla \psi$, $ en\,\Omega \nabla T_{\perp}$ and $\nabla\bdot{\sf P}_{CGL}$ in Eqs.~(\ref{eq:parkin_ex})-(\ref{eq:field_ex}), on the other hand, describe standard energy-exchange processes.

\subsection{Comparison with Previous Models}

We now write explicit final expressions for the gyrofluid density $n$, the gyrofluid parallel velocity $u_{\|}$, and the parallel and perpendicular gyrofluid pressures $p_{\|}$ and $p_{\bot}$. The gyrofluid continuity is expressed in expanded form as
\begin{eqnarray}
\frac{d_{E}n}{dt} \;+\; \nabla\bdot\left[ n \left( u_{\parallel}\;\bhat^{*} + {\bf u}_{C} + {\bf w}_{C}\right) \right] 
 & = & \left[\; \K(p_{\perp}) \;+\; n\;\K(e\,\psi) \;+\; T_{\bot}\;\K(en\,\Omega) \;\right] \nonumber \\
 &  &\mbox{}+\; {\bf u}_{D}\bdot\nabla(en\,\Omega) \;+\; {\bf w}_{\Omega}\bdot\nabla n,
\label{eq:densitytot}
\end{eqnarray}
where $d_{E}/dt \equiv \partial/\partial t + {\bf u}_{\psi}\bdot\nabla$. With the insertion of the 
diamagnetic-cancellation term (\ref{eq:diacan_upar}), the evolution equation for the gyrofluid parallel velocity is expressed in expanded form as
\begin{eqnarray}
mn\,\left( \pd{u_{\|}}{t} \;+\; {\bf u}^{\prime}\bdot\nabla u_{\|} \right) \;-\; p_{\bot}\,\K(mu_{\|}) & = & 
-\;\bhat^{*}\bdot\left[\; en\;\nabla\psi \;+\; \nabla p_{\bot} \;+\; T_{\bot}\;\nabla(en\Omega) \;\right] \nonumber \\
 &  &\mbox{}-\; \nabla\bdot\left[\; (p_{\Delta} - e\Omega\,p_{\bot})\;\bhat\;\right],
\label{eq:momentumtot}
\end{eqnarray}
where ${\bf u}^{\prime} \equiv {\bf u} - {\bf u}_{D}$ denotes the gyrofluid velocity without its diamagnetic contribution. Lastly, with the insertion of the diamagnetic-cancellation terms (\ref{eq:addpar})-(\ref{eq:addperp}) and 
(\ref{eq:Q_parperp}), the gyrofluid parallel and perpendicular pressure equations are
\begin{eqnarray}
\pd{p_{\|}}{t} & = & -\;\nabla\bdot\left( p_{\|}\;{\bf u} \;+\; 2\,p_{\parallel}\,u_{\|}\;\bhat \;+\; 2\,{\bf q}_{\|\bot}
\right) \;+\; 2\,{\bf u}\bdot\nabla\bdot\left( p_{\|}\,\bhat\bhat\right) \nonumber \\
 &  &\mbox{}+\; 4\,(1 + e\Omega)\,{\bf q}_{\parallel \perp}\bdot (\bhat\bdot \nabla\bhat), 
\label{eq:ppar_tot}
\end{eqnarray}
and
\begin{eqnarray}
\pd{p_{\bot}}{t} & = & -\;\nabla\bdot\left[\; p_{\bot}\;{\bf u} \;+\; p_{\perp}\;({\bf I} - \bhat\bhat)\bdot{\bf u} 
\;+\; (1 + e\Omega)\,{\bf q}_{\perp\perp}\;\right] \;+\; {\bf u}\bdot\nabla\bdot\left[\; p_{\perp}\;
({\bf I} - \bhat\bhat) \;\right] \nonumber \\
 &  &\mbox{}-\; 2\,{\bf q}_{\parallel \perp}\bdot (\bhat\bdot \nabla\bhat) \;-\; {\bf q}_{\bot\bot}\bdot\nabla(e\Omega), 
\label{eq:pper_tot}
\end{eqnarray}
where the heat fluxes ${\bf q}_{\|\bot}$ and ${\bf q}_{\bot\bot}$ are defined in Eqs.~(\ref{eq:qpar}) and (\ref{eq:qperp}).

The results of the present two-temperature gyrofluid model can be compared with the previous one-temperature model 
presented in Ref.~\cite{ours}, where the gyrofluid equations are derived by including the perpendicular temperature only. Since the FLR-corrected CGL pressure tensor (\ref{eq:CGL_mod}) includes the parallel pressure $p_{\|}$, the two gyrofluid models agree except for the terms that arise from the pressure anisotropy $p_{\Delta} = p_{\|} - p_{\bot}$.

Our results can also be compared with the gyrofluid model of Beer and Hammett \cite{beer}, since this Beer-Hammett model is the most extended one, which includes closures and contains all the previously developed gyrofluid models. The equations of evolution of the gyrofluid moments presented here are nearly identical to those of Beer and Hammett. The differences that arise can be separated in two categories: (1) differences in the non-FLR terms and (2) differences in the FLR terms.

The non-FLR terms coincide almost exactly in the two models, except terms that come from closures of higher-order moments, which our model is not able to retrieve. In our model, gyrofluid moments higher than pressure moments cannot be included in the gyrofluid Lagrangian, and the closures are done automatically when choosing the constraints for the variation. Thus, for instance, Landau damping is not included in our model, though it can be added by hand afterwards as long as the energy conservation remains exact. A special difference in the non-FLR terms between the two models - and the only one of the kind that appears - is a different term in the momentum equation (\ref{eq:momentumtot}): the magnetic term
$\K(u_{\parallel})$ in our model compared to $2\,\K(u_{\parallel})$ in Ref.~\cite{beer}. This difference arises from the fact that the additional magnetic contribution originates from the parallel-parallel heat flux 
${\bf q}_{\parallel \parallel}$, which we cannot fully retrieve.

The differences in the FLR terms between the two models arise from the higher-order moment closures, but also because of the energy conservation law. Especially in the FLR terms, previous models do not conserve energy. Although our polarization equation reduces to the local one \cite{scott} and so is in agreement with the local model in the local limit, we do not exactly retrieve the results of Ref.~\cite{scott}. That is, although we find a strong correlation between $n$ and $\|\phi\|$ and $p_{\perp}$ and $\Omega$, there exists another FLR term that is not correlated with 
$p_{\perp}$. This difference arises because of the constraints introduced here, which lead to another closure than that of Ref.~\cite{scott}.  

\section{Summary and conclusions}

In this paper we have derived a set of electrostatic gyrofluid equations for an anisotropic plasma, that describe the evolution of density, momentum, parallel and perpendicular pressures. The fully inhomogeneous four-moment model also includes a polarization equation, from which the electric field is computed, and satisfies an exact energy conservation law that includes energy-exchange terms involving parallel kinetic energy, internal energy, and field energy.

To guarantee energy conservation, we have used the Lagrangian approach in which all the equations are derived by a variational principle. Diamagnetic cancelations were taken into account in a second step, by the addition of terms that include gyrofluid moments higher than the pressure. All non-electrostatic diamagnetic-corrections terms were computed from the Vlasov equation, and FLR-corrected terms associated with $\phi$ were chosen so that the energy conservation law is still satisfied.

The validity of the Lagrangian and the constraints introduced for the variational procedure was verified by the validity of the final evolution equations and the energy conservation theorem. The fact that the set of equations conserves energy makes the model suited for turbulence computations, and especially for those that treat large-amplitude disturbances, or those that treat strong variations of the plasma parameters. Moreover, since the set of equations is fully inhomogeneous, it is suitable for studying the nonlinear evolution of various fields, independent of the linear growth phase.

The successful application of this technique to the four-moment
gyrofluid model show that it is feasible, although not trivial, to
extend the Lagrangian formulation to include also the heat fluxes as
dynamical variables, and, thus, arrive at a complete six-moment
model. This extension will be treated in future publications.

\vspace*{0.2in}

\noindent
{\large {\bf Acknowledgments}}

We gratefully acknowledge useful discussions with Profs.~K.~Lackner, D.~Pfirsch, and H.~Weitzner, as well as Dr.~D.~Correa-Restrepo. One of us (AJB) also wishes to thank Profs.~K.~Lackner and S.~G\"{u}nter for their invitation to visit the Max-Planck-Institut f\"{u}r Plasmaphysik in Garching.

\newpage

\end{document}